\documentclass[11pt]{article}
 
\usepackage{amsmath,amsthm}
\usepackage{amsthm}
\usepackage{amssymb}
\usepackage{amsfonts} 
\usepackage{color}

\usepackage[latin1]{inputenc}

\usepackage{graphicx}

\newcommand{\sect}[1]{\setcounter{equation}{0}\section{#1}}
\newcommand{\subsect}[1]{\subsection{#1}}

\def\1{\'{\i}}

\begin{document}

\bigskip
 
\begin{center}
{\Large{\bf{The Volterra Integrable case}}}

\bigskip 
\bigskip

\begin{center}
M.Scalia$^a$ and O. Ragnisco$^b$
\end{center}

\noindent
$^a$Dipartimento di Matematica, Universit\'a degli Studi ``La Sapienza"
\\P.le A.Moro 2, I-00185 Roma


\noindent
$^b$Istituto Nazionale di Fisica Nucleare, Sezione di Roma 3,
\\Via della Vasca Navale 84,00146 Roma

\end{center}

\bigskip

\begin{abstract}
\noindent 
In this short note we reconsider the integrable case  of the Hamiltonian $N$-species Volterra system, as it has been introduced by Vito Volterra in 1937. In the first part, we discuss the corresponding conserved quantities, and comment about the solutions of the equations of motion. In the second part we focus our attention on the properties of the simplest model, in particular on period and frequencies of the periodic orbits. The discussion and the results presented here are to be viewed as a complement to a more general work, devoted to the construction of {\it a global stationary state model for a sustainable economy in the Hamiltonian formalism}.
\end{abstract}

\bigskip\bigskip 

\noindent
PACS:\quad  
  

\noindent
KEYWORDS:  Generalized Volterra Model, Hamiltonian formulation, Integrable Systems

\sect{Introduction}
The current twofold crisis, at a global level, of Economics and
Environment has been deeply investigated and reported by many
authors, some of which have deeply criticized the deafness of
Economists with respect to the environmental crisis, that at most
has been assessed for its dramatic consequences on GDP (see the
well-known ``Stern's Report" \cite{Stern}. The double crisis and its
entanglement would request models, even before than a theory,
able to put together economic and environmental variables in
order to build a global stationary state to rule present predicament
in the perspective of a sustainable scenario. The latter theme is not
a news, several attempts having been realized starting from the
Seventies for a definition of ``steady state"(\cite{Georgescu 1}, \cite{Georgescu 2}, \cite{Daly}) but it could be
useful to face the problem with other scientific tools, as it has been
recently proposed in \cite{Scalia1}, \cite{Scalia 2}, \cite{Scalia 3}, where the leading idea is to put together
pairs, each constituted by an economic variable and an
environmental one, that present a behavior ``predator-prey" type,
as it is suggested by some of the most important pairs one can
select for the model. Already fifty years ago, a similar idea was
applied to build the Goodwin model, but with a pair of variables
only economic to describe an economic cycle \cite{Goodwin}, the so called
``class struggle" model, where a variable linked to the wage rate
assumes the role of predator and the one giving employment rate
is the prey.

\sect{The Model}

As is well known, the original idea by Vito Volterra \cite{Volterra1} was that of
determining the evolution of a two species biological system, the
so-called ``predator-prey" model, answering a question raised by
his  son in law, the biologist Umberto
d'Ancona \cite{dancona}, who was wondering why the total catch of selachians
(mostly sharks) was considerably raising during World War 1,
with respect to other more desirable kind of fishes, in
correspondence with the decrease of fishing activity \cite{Brown}. To answer
that question, Vito Volterra constructed a dynamical system that
enabled him to identify the essential features of what was going on,
elucidating the properties entailing the existence of a stable
equilibrium configuration (and of periodic orbits in its
neighbourhood), and unveiling the asymptotic behaviour of the
system under general initial conditions. He quickly realized that
the predator-prey model was just the simplest example in a large
class of biological, or rather {\it ecological} systems with pairwise
interaction. He was soon interested in understanding the
mathematical properties of the $N$ species pairwise interacting
model, and devoted a considerable effort to find suitable
Lagrangian and Hamiltonian formulations, with the final aim of
achieving a description where the deep analogy with the well
established theory of mechanical systems stemming from the
Maupertuis {\it minimal action} principle be made transparent. We
would say that not the whole ``Biological-Mechanical" dictionary
that he proposed in his famous paper (dating back to 1937),
{\it Principes de Biologie Mathematique} \cite{Volterra2}, resisted the future
developments of both disciplines, and some of the notions he tried
to introduce look nowadays a bit artificial. But we believe that the
core of his  derivation is still alive, as it has been witnessed by a
very wide spread applications over about a century in many
scientific research subjects, such as Populations demography, Bio-physics, Biomedicine,
Ecology and also Economics. We notice that in \cite{Volterra2} his main aim was the formulation of this generalized model in  a Hamiltonian language, with the purpose of elucidating  the algebraic conditions leading to a completely integrable model. So we think that it could be worth recalling the key
ingredients of Volterra's approach, and even emphasizing the role
that the very special case of completely integrable dynamics could
play in the search for Stationary State models in the economic-ecological
framework. His general Predator-Prey model reads:

\begin{equation}
\frac{d N_r}{dt} = \epsilon_r N_r + \sum_{s\ne r=1}^N A_{rs}N_rN_s \label{Volt1}
\end{equation}
In (\ref{Volt1}), we have set all the parameters  introduced in \cite{Volterra1} $\beta_r =1 \forall r$; $\epsilon_r$ are the natural growth coefficients of each species and $A_{rs} $ are interaction coefficients between species $r$ and species $s$ that account for the probability 
 of encountering between two individuals.
 Moreover, for the moment we assume that the matrix $A$, whose elements are $A_{rs}$, is nonsingular. 
 The last requirement ensures that the system of equations defining the equilibrium configurations, namely
 \begin{equation}
0= \epsilon_r + \sum_{s=1}^N A_{rs}N^s \label{equi}
\end{equation}
has a unique solution, say $Q_r$, $r=1,\cdots,N$. If, in addition, according to \cite{Volterra1}, we require $A$ to be skew-symmetric, $N$ has to be even, and the eigenvalues of $A$ must be purely imaginary and complex conjugate in pairs.

\subsect{Lagrangian and Hamiltonian formulation}

As many other researchers of his time, Volterra was feeling more assured if a phenomenon quantified by Mathematics could find an analogue with Mechanics, that moreover allowed resorting to the powerful formalism and theorems of the latter. The quantity of life of the species $r$, defined as $X_r = \int_0^t N_r(t^\prime)dt^\prime$ suggested to Volterra the introduction of a biological, or rather $ecological$, Lagrangian $\Phi$, defined as the sum of three terms

\begin{equation}
\Phi = X+ \frac{1}{2}Z +P \label{lagr1}
\end{equation}
where
\begin{equation}
X= \sum_r  X_r^\prime \log  X_r^\prime;   ~ Z= \sum_{rs}A_{rs}X_r^\prime X_s; P = \sum_r \epsilon_r X_r\label{lagr2}
\end{equation}
In terms of (\ref{lagr1}), (\ref{Volt1}) can be written as Euler-Lagrange equations

\begin{equation}
\frac{d}{dt}\frac{\partial \Phi }{\partial X_r^\prime} - \frac{\partial \Phi}{\partial X_r} = 0 \label{EL}
\end{equation}
	
\noindent
yielding the ODEs
\begin{equation}
X_r^{''}=(\epsilon_r +\sum_s A_{sr}X_s^\prime)X_r^\prime\label{ODEs}
\end{equation}
\noindent
which are just (\ref{Volt1}), up to the substitution $N_r=X_r^\prime$, the $\prime$ denoting time-derivative. 

\subsect{From Lagrange to Hamilton}

\noindent
The linear momenta, canonically conjugated to the quantities of life, are defined as
\begin{equation} 
p_r =\frac{\partial \Phi}{\partial X_r^\prime}= \log X_r^\prime+1 +\frac{1}{2}\sum_s A_{rs}X_s\label{momenta}
\end{equation}

\noindent
whence
\begin{equation}
X_r^\prime= \exp(p_r -1-\frac{1}{2}\sum_s A_{rs}X_s)
\end{equation}
\noindent
Through a transformation of Legendre type we define the Hamiltonian 
\begin{equation}
{\mathcal H} = \Phi - \sum_r X_r^\prime p_r=\sum_r\epsilon_rX_r-X_r^\prime =\sum_r\epsilon_rX_r - \exp(p_r -1-\frac{1}{2}\sum_s A_{rs}X_s)\label{Ham}
\end{equation}
Volterra \cite{Volterra2} showed that (\ref{Volt1}) can be written in the standard hamiltonian form 
\begin{equation}
X_r^\prime = -\frac{\partial {\mathcal H} }{\partial p_r}\label{Ham1}
\end{equation}
\begin{equation}
p_r^\prime =\frac{\partial {\mathcal H} }{\partial X_r}\label{Ham2}
\end{equation}
As the derivation is a bit tricky we prefer to sketch it.

\noindent
One starts by writing

$$\frac{\partial {\mathcal H} }{\partial X_r}=\epsilon_r -\frac{\partial}{\partial X_r}[\sum_k\exp(p_k-1-\frac{1}{2}\sum_s A_{ks}X_s)] =$$

\noindent
$$\epsilon_r+\frac{1}{2}\sum_kA_{kr} \exp(p_k-1-\frac{1}{2}\sum_s A_{ks}X_s)=\epsilon_r+\frac{1}{2}\sum_kA_{kr}X_k^\prime=$$

\noindent
$$\frac{\partial \Phi}{\partial X_r} = \frac{d}{dt}\frac{\partial \Phi}{\partial X_r^\prime} = p_r^\prime$$

\noindent
Notice however that (\ref{Ham2}) holds just if the Euler-Lagrange equations (\ref{EL}) are satisfied. It is easily seen (see again [4]) that the above Hamiltonian system has the following $N$ independent non autonomous integrals of motion:

\begin{equation}
{\mathcal H_r} = \frac{p_r + \frac{1}{2}\sum_s A_{rs}X_s}{\epsilon_r} -t ~~r=1,\cdots,N. \label{H_r}
\end{equation}

\noindent
whence  one can select $N-1$ autonomous integrals by taking for instance ${\mathcal H_{1,r}}\equiv {\mathcal H_r}-{\mathcal H_1}$, and have a complete set by adding the Volterra $N-$ species Hamiltonian (\ref{Ham}). 

\noindent
A more modern approach to the Hamiltonian structure underlying the generalised Volterra system can be be found  in an elegant paper by R.Loja and W.Oliva \cite{LO}, where a Poisson morphism is established between  the original system, living in ${\mathbb R}^N$ and equipped with a quadratic Poisson structure, and  the one recast, after Volterra, in a canonical Hamiltonian form  and thus living in   ${\mathbb R}^{2N}$.

\noindent
However,  in our opinion, a crucial question to ask is whether there exists a special form 	of the matrix elements $A_{rs}$ that entails involutivity of that complete set of integrals of motion. It turns out that this form has been found by Volterra himself \cite{Volterra2} and is the following:

\begin{equation}
A_{rs} =\epsilon_e \epsilon_s (B_r-B_s)~~~r,s=1,\cdots N	\label{invo}	
\end{equation}			

\noindent
where $N$ is the number of competing populations and the $B_r$ are real (and positive) numbers. 
The matrix $A$ can be written as:
\begin{equation}
 A = [B,\epsilon \otimes \epsilon] \label{A}
\end{equation}

\noindent
where $B$ $=diag(B_1,\cdots,B_N)$, and $\epsilon$ is the vector $(\epsilon_1,\cdots,\epsilon_N)^t$, meaning that $ A$ is the commutator of a diagonal matrix with distinct entries and a rank one matrix.  
Coming back to the Lagrangian formulation, namely rewriting:
$$X_r^\prime = \exp(p_r -1- \frac{1}{2}\sum_s A_{rs}X_s)$$

\noindent
or

$$p_r=  \frac{1}{2}\sum_s A_{rs}X_s +1+\log X_r^\prime$$

\noindent
We see that (2.12) can be rewritten as

\begin{equation*}
{\mathcal H_r} =\frac{ \log X_r^\prime + \sum_s A_{rs}X_s}{\epsilon_r}- t~~r=1,\cdots,N. \label{H_r}
\end{equation*}

\noindent
whence

 \begin{equation*}
 \exp(\epsilon_r{\mathcal H_r}) =X_r^\prime\exp(\sum_s A_{rs}X_s -\epsilon_r t )
 \end{equation*}
 
 \noindent
 that in the integrable case read
 
  \begin{equation*}
 \exp(\epsilon_r{\mathcal H_r}) =X_r^\prime\exp(\epsilon_r(B_r \tilde X - <{\vec\epsilon}|B|X>  - t )
 \end{equation*}
 
 \noindent
 where we have set
 $$\tilde X = \sum_s \epsilon_s X_s$$
 
 \noindent

\medskip 
 \subsection{A note about the equilibrium conditions}

 \noindent
 As a matter of fact, the highly degenerate structure of $ A$ in the integrable case entails that its kernel consists of all vectors which are orthogonal  to both $|{\vec \epsilon}>$ and 
 $B|{\vec \epsilon}>$, which are certainly linearly independent as by assumption the entries of the matrix $B$ are all distinct. So, its Kernel $Ker (A)$ is a linear space of codimension 2;  by ``Rouch\'e -Capelli", its Range $R(A)$ is two-dimensional, and {\it the equilibrium configuration is highly non-unique}. 
 
 \noindent
 Indeed, in the integrable case the equilibrium condition reads:
 
 $$\sum_{s=1}^N\epsilon_s(B_s-B_r)Q_s =1$$
 
 \noindent
 where we have denoted by ${Q}_s (s=1,\cdots,N)$, the equilibrium configuration. It follows (being $\sum_{s=1}^N \epsilon_s Q_s =0$) that they belong to the hyperplane:
 
 \begin{equation}
 \sum_{s=1}^N\epsilon_sB_sQ_s =1\label{Hyper}
 \end{equation}

\noindent
So, the equilibrium configurations are defined as the intersection of the two hyperplanes 

\noindent
$$<\epsilon|Q>=0;\quad <B|\epsilon|Q>=1$$

\noindent
Incidentally, we notice that it does not seem that Volterra had paid a special attention to the singular nature of the matrix $A$ in the integrable case for any $N$ larger than $2$.

\noindent
 So far, we have looked at the equilibrium in terms of the natural coordinates. The situation is unfortunately much less clear in the canonical setting. First of all, rephrased in terms of the quantities of life, the above equations take the form:
 
 $$0=\epsilon_r + \sum_{s=1}^N A_{rs}X_s^\prime.$$
 
 \noindent
 which is a linear system in the variables $X_s^\prime$.
 \noindent
 In turn. by expressing $X_s^\prime$ in terms of the canonical variables we get:
 
 \begin{equation}
 0=\epsilon_r + \sum_{s=1}^N A_{rs}\exp[p_s-1-\frac{1}{2}\sum_kA_{sk}X_k]\label{equican}
 \end{equation}
 
 \noindent
 On the other hand from the Hamilton's equations we get:
 
 \begin{equation}
 p^\prime=0=-\frac{\partial {\mathcal H} }{\partial X_r }=\epsilon_r +\frac{1}{2}\sum_{s=1}^N A_{sr}X_s^\prime=\epsilon_r  +\frac{1}{2}\sum_{s=1}^N A_{sr}\exp(p_s-1-\frac{1}{2}\sum_{q=1}^N A_{sq}X_q)
 \end{equation}
 
 \noindent
 Evidently formulas (2.16) and (2.17) do not match, because of the factor $\frac{1}{2}$ in front of the exponential in (2.17),  which in (2.16) is missing. This discrepancy has to be understood.
 
 \noindent
 
 \subsection{Back to the integrable case}
 \noindent
 However, once realized that integrability implies the non-uniqueness of the equilbrium configuration ($\forall$ $N>2$), what is really crucial to understand is whether, for the special form of the matrix $A$ given by (\ref{invo}) the integrals of motion are still independent. But a glance at (2.12) shows that the linear dependence of those integrals upon the momenta $p_r$ is in no-way affected by by the specific form of the matrix $A$ (while the requirement that the $B_r$ be all distinct is mandatory!), so that the rank of the Jacobian matrix constructed with the gradients of the integrals of motion with respect to the canonical coordinates is maximal (namely $N$) whatever be that form. So, the integrable version of the $N$-species Volterra system is again a genuine hamiltonian system with $N$ degrees of freedom. 
 
 \noindent
 Here we write down explicitly the expression of the Hamiltonian and of the integrals of motion in the integrable case. 
 
 \begin{equation}
 {\mathcal H}_{int} = \sum_{r=1}^N \epsilon_r X_r-\exp[p_r-(\epsilon_r/2)\sum_{s=1}^N\epsilon_s(B_r-B_s)X_s]\label{H_int}
 \end{equation}
 
 \noindent
 and
 \begin{equation}
{\mathcal H}_r = p_r/\epsilon_r + (1/2)\sum_{s=1}^N\epsilon_s(B_r-B_s)X_s -t \quad r=1,\cdots,N\label{Hr_int}
\end{equation}

\noindent
so that
\begin{equation}
{\mathcal H}_{rl}\equiv {\mathcal H}_r -{\mathcal H}_l =p_r/\epsilon_r -p_l/\epsilon_l +\frac{1}{2}(B_r-B_l)\sum_{s=1}^N \epsilon_s X_s, \quad s=1,\cdots, N\label{Hrl}
\end{equation}

\noindent
The constants of motion (\ref{Hrl})are mutually in involution. So we can take for instance $l=1$ and get $N-1$ independent integrals of motion in involution. The set can be completed by adding any function of the Hamiltonian. for instance the Hamiltonian  itself.

\noindent
The above formulas clearly show that, in the integrable case, both the Volterra Hamiltonian and the involutive family of integrals of motion depend on the quantities of life only through the
 inner  products $<{\vec \epsilon}|X>$ and $<{\vec \epsilon}|B|X>$. Of course, if one comes back to the Lagrangian formulation, one will get  expressions involving the variables $X_s$ and $X_s^\prime$. Let us also notice that
 
 \begin{equation}
 \exp(\epsilon_r {\mathcal H}_r )= \exp[p_r + \epsilon_r/2\sum_{s=1}^N \epsilon_s(B_r-B_s)X_s -\epsilon_r t]\label{exp}
 \end{equation}

 \noindent
 Obviously, one could choose $\exp({\mathcal H_{rl}})$ as an alternative legitimate form for an involutive family of integrals of motion. In the simplest nontrivial case, $N=2$, (\ref{H_int}) reads:
 
 \begin{equation}
 {\mathcal H}_v = \epsilon_1X_1 + \epsilon_2 X_2 - \exp[p_1-(1/2)\epsilon_1\epsilon_2(B_1-B_2)X_2] - \exp[p_2 + (1/2)\epsilon_1\epsilon_2(B_1-B_2)X_1]\label{Hv}
 \end{equation}
 
 \noindent
 The above formula can be slightly simplified through the canonical transformation (in fact, a rescaling):
 \begin{equation}
 p_j  \to \tilde p_j = p_j/\epsilon_j; \quad X_j\to \tilde X_j = \epsilon_jX_j
 \end{equation}
 that maps (\ref{Hrl}) into:
 \begin{equation}
 {\mathcal H}_{rl} =\tilde p_r -\tilde p_l +(1/2)(B_r-B_l)<1|\tilde X>
 \end{equation}
 
 \noindent
 and (\ref{Hv}) into:
 \begin{equation}
 { \mathcal H}_v = \tilde X_1 + \tilde X_2 - \exp[\epsilon_1(\tilde p_1-(1/2)(B_1-B_2)\tilde X_2)] - \exp[\epsilon_2(\tilde p_2 + (1/2)(B_1-B_2)\tilde X_1]\label{newHv}
 \end{equation}
  
  \noindent
  For an arbitrary $N$, the same canonical transformation yields:
  \begin{equation}
   {\mathcal H}_{int} = \sum_{r=1}^N \tilde X_r-\exp[\epsilon_r(\tilde p_r- 1/2 \sum_{s=1}^N(B_r-B_s)\tilde X_s)]\label{newH_int}
 \end{equation}

\noindent
Always sticking on the integrable case, it looks a bit surprising that the {\it reduction to quadratures} was not  performed by Volterra in the canonical setting, but rather in terms of the natural coordinates, namely those denoting the population numbers of the different species. In fact, Volterra defines the quantities ${\mathcal N} := \sum_s \epsilon_sN_s$ and ${\mathcal M}:=1-\sum_s \epsilon_s B_s N_s$, then rewriting the original dynamical system (\ref{Volt1}) as:
 \begin{equation}
 \dot N_r = \epsilon_r N_r( 1+\sum_s(B_r-B_s)\epsilon_sN_s)
 \end{equation}
 
 \noindent
 or, in other terms:
  \begin{equation}
 \dot N_r = \epsilon_r N_r( B_r {\mathcal N} + {\mathcal M})
 \end{equation}
  \noindent
  namely:
   \begin{equation}
   (1/\epsilon_r) d/dt \log N_r = ( B_r {\mathcal N} + {\mathcal M})
   \end{equation}
   Note that the previous equation implies $\sum_r(\dot N_r - \epsilon_rN_r) =0$. In terms of the quantities of life $X_r$ we have $\sum_r (X_r ^{\prime \prime} -\epsilon_r X_r^\prime) =0$, yielding the conserved quantity $C_0=\sum_r ( \epsilon_r X_r - X_r ^\prime)$, which is just the Hamiltonian (\ref{Ham}). It is evident that the involutivity constraints on the coefficients $A_{rs}$ entail a typical ${\it Mean~Field}$ dynamics. Each species interacts with the others through the ${\it collective~variables}$ ${\mathcal N}$ and ${\mathcal M}$. By taking two different values of the index $r$ and subtracting, the variable $ {\mathcal M}$ can be eliminated, resorting to:
  \begin{equation}
  \frac{(1/\epsilon_r) d/dt \log N_r -(1/\epsilon_s) d/dt \log N_s}{B_r-B_s} = {\mathcal N} 
  \end{equation}
  
  \noindent
  By setting $F_k = N_k^{(1/\epsilon_k)}$, we see that the above formula implies that the time derivative of the differences $g_k-g_l$, where $g_m:=\frac{\log (F_1/F_m)}{B_1-B_m}$ vanishes. So, we immediately get $N-2$ integrals of motion written in terms of the natural variables. 
 
 \noindent
The remaining integrals can be obtained by taking the Hamiltonian itself and the quantity $L:=\sum_{r=1}^N \epsilon_rQ_r{\mathcal H}_r$. 

\noindent
Note, however, that as the $Q_r$ are not uniquely defined, this non-uniqueness affects $L$ as well. Of course such ``superabundance" of integrals of motion does not really take place, because the different $L$ will not be functionally independent. It will be enough to consider a single representative element of the class of equilibrium configurations. 

\medskip
\subsection{The Hamiltonian (\ref{newHv}) is indeed integrable!}
Let us slightly simplify the notations, by setting $\mu := B_2-B_1$, and introducing the new {\it{canonical variables}}:

\begin{equation}
P_1 = \frac{1}{\sqrt \mu}(\tilde p_1 +\frac{1}{2} \tilde X_2); ~~~Q_1 = \frac{1}{\sqrt \mu} (-\tilde p_2 + \frac{1}{2} \tilde X_1)
\end{equation}

\begin{equation}
P_2 = \frac{1}{\sqrt \mu}(\tilde p_1 -\frac{1}{2} \tilde X_2); ~~~Q_2 = \frac{1}{\sqrt \mu} (\tilde p_2 + \frac{1}{2} \tilde X_1)
\end{equation}

\noindent
In terms of these new  variables, the first integral:
\begin{equation}
{\mathcal H}_{12} =\tilde p_1 -\tilde p_2 +(1/2)(B_1-B_2)<1|\tilde X>
\end{equation}
takes the form
\begin{equation}
{\mathcal H}_{12} = {\sqrt \mu}(P_2-Q_2)
\end{equation}

\noindent
while the two-particle Hamiltonian reads:

\begin{equation}
{\mathcal H}_v = \frac{1}{\sqrt \mu}[Q_1 + Q_2 +P_1-P_2] -\exp (\epsilon_1 P_1) -\exp(\epsilon_2 Q_1)
\end{equation}

\noindent
Inserting the first integral (2.34), on the level surface ${\mathcal H}_{12}= C$, up to an irrelevant additive constant we can finally write:

\begin{equation}
{\mathcal H}_v = \frac{1}{\sqrt \mu}(Q_1+P_1) -\exp (\epsilon_1 P_1) -\exp(\epsilon_2 Q_1)
 \end{equation}

\noindent
It follows that, in terms of these new coordinates, the  Hamiltonian  (2.25) becomes a ``one-body" hamiltonian (integrable by definition), which is nothing but the traditional Volterra-Lotka Hamiltonian.

\subsection{Suggesting a mixture}
 An interesting attempt to get a more realistic system could be performed  by taking the Hamiltonian to be a linear combination of the sum of two non interacting predator-prey models, where prey and predator are  taken as a conjugate symplectic pair, and the Hamiltonian version of the $N~=~2$ original Volterra system.
 
\noindent
Accordingly we   will start with a non-interacting two-body system  ({\it in the economic-echological interpretation}   each {\it body} is in fact a pair of conjugated quantities)     described by the sum of two independent Hamiltonians ${\mathcal  H}_1$ and ${\mathcal H}_2$.
 
 \begin{equation}
 {\mathcal H}_0 ={\mathcal H}_1 + {\mathcal H}_2 = \sum_{i=1}^2 \epsilon_i q_i + \eta_1 p_i - \alpha_i\exp(q_i) -\beta_i \exp (p_i) \label{H0}       
 \end{equation}      

\noindent 
 The hamiltonian (\ref{H0}) has a single equilibrium   configuration corresponding to the point in ${\mathbb R}^4$ $P= (\log(\eta_1/\beta_1),\log(\epsilon_1/\alpha_1),\log(\eta_2/\beta_2),\log(\epsilon_2/\alpha_2)$. Of course the positivity requirements are fulfilled provided that $\epsilon_i/\alpha_i,\eta_i/\beta_i >1 ~ (i=1,2)$ . Under the above hypotheses $P$ will e a center, and the phase space will be foliated by two-dimensional tori, parametrized by the energy $E_1$ and $E_2$.  We expect that the family of plane  curves $f(x,y)=C$, $C$ being a positive constant, with 
 \begin{equation}
 f(x,y) := \alpha \exp(x) + \beta \exp(y) -\epsilon x - \eta y \label{integralcurve}
 \end{equation}

\noindent 
 be bounded and contained in the first quadrant of the $(x,y)$ plane whenever $\epsilon$ and $\eta$ are both positive but smaller than $\alpha$ and $\beta$ respectively. 

\medskip
\subsection{Linear stability and small oscillations for the decoupled system}                                                                                                                                                                                                                                                                                                                                                                                                                                                                                                                                                                                                                                                                                                                                                                                                                                                                                                                                                                                                                                                                                                                                                                                                                                                                                                                                                                                                                                                                                                                                                                                                                                                                                                                                                                                                                                                    
    
The equations of motion for the Hamiltonian (\ref{H0}) read:

\begin{equation}
\dot q_i =\eta_i - \beta_i \exp(p_i)\quad (i=1,2)
\end{equation}
\begin{equation}
\dot p_i =-\epsilon_i + \alpha_i\exp(q_i) \quad (i=1,2)
\end{equation}

\noindent
In terms of exponential functions, the equilibrium configuration is  given by

\begin{equation*}
\exp(\tilde q_i)=\epsilon_i/\alpha_i; \quad \exp(\tilde p_i)= \eta_i/\beta_i
\end{equation*}

\noindent
However, it is worth noticing that in the natural variables (those originally introduced by Volterra)a discussion about equilibrium stability and small oscillations will be much easier.
Accordingly, we set $x_i= \exp(q_i), \quad y_i= \exp(p_i)$, so that the Hamilton's equations take the form: 
\begin{equation}
\dot x_i=-\beta_i (y_i-\tilde y_i) \quad \dot y_i = \alpha_i(x_i-\tilde x_i)
\end{equation}

\noindent
Introducing the differences:

$$ \xi_i =x_i-\tilde x_i; \quad \zeta_i= y_i-\tilde y_i$$

\noindent
which are clearly $O(\epsilon)$ so that  products like $ \xi_i  \zeta_i$ will be $O(\epsilon^2)$,
we get:
\begin{equation}
\dot \xi =-\beta_i\tilde x_i\zeta_i +O(\epsilon), \quad \dot \zeta_i = \alpha_i \tilde y_i \xi_i +O(\epsilon)
\end{equation}

\noindent
Consequently, the secular equation will 
be the following biquadratic equation:   

$$ \lambda^4 + \lambda^2 (\alpha_1\beta_1\tilde x_1 \tilde y_1 + \alpha_2\beta_2\tilde x_2 \tilde y_2) +\alpha_1\beta_1\tilde x_1 \tilde y_1\alpha_2\beta_2\tilde x_2 \tilde y_2 =0$$

\noindent
whose solutions are:

$$\lambda_i^2 = -\alpha_i\beta_i\tilde x_i \tilde y_i$$

\noindent
Inserting the explicit formulas for the equilibrium positions, we get:

$$\lambda_i^2=- \epsilon_i \eta_i$$

\noindent
meaning that, assuming the products  $\epsilon_i \eta_i$ to be all positive, the eigenvalues will be purely imaginary as expected. Of course, being decoupled from the very beginning, the system will keep its decoupled nature in the small oscillation regime as well. Investigating the small oscillation regime for the simplest coupled Volterra model  will be of course more interesting. But a deeper look at the equations of motion of such model, possibly not written in Hamiltonian form, will be needed to identify its equilibrium positions. 

\noindent
The equation for the integral curve (conservation of energy) (\ref{integralcurve}) can be written in a slightly more elegant way. Indeed, introducing the equiibrium position and conveniently rescaling the independent variables, $f(x,y)=C$  acquires the form:

\begin{equation}
\epsilon f(\xi) + \eta f(\zeta) =C- (\epsilon + \eta) := k^2 \label{curve}
\end{equation}

\noindent
having set 
\begin{equation}
f(x)= \exp(x)-x-1
\end{equation}

\noindent
we remark that the function $f(x)$ defined by (2.44) is nonnegative on the whole real line, reaches its minimum at the origin where it vanishes, and is monotonically decreasing (resp. increasing)
in the negative (resp. positive) semiaxis. Therefore there is a diffeomorphism mapping the curve $\epsilon f(\xi) + \eta f(\zeta) = k^2 (\epsilon >0, \eta>0)$ to the curve $aX^2+bY^2=1$ for some $a>0,b>0$.

One way to couple the two systems could be by adding to (\ref{H0}), through a dimensionless parameter $\lambda$, the generalized Volterra predator-prey hamiltonian (\ref{Hv}), written in Volterra's hamiltonian variables, in the simplest nontrivial case ($N =2$).  Both (\ref{H0}) and (\ref{Hv}) are completely integrable, but since they are not in involution the coupled system will lose the complete integrability property. However by acting on the control parameter $\lambda$ one may think of following the behaviour of such non-integrable system, and see the corresponding merging of the two tori in a single 4-dimensional surface. The suggested Hamiltonian reads:
$${\mathcal H}_\lambda = {\mathcal H}_0 + \lambda {\mathcal H}_v$$
where of course ${\mathcal H}_v$ is given by (\ref{Hv}).
However, It would be reasonable to require that the coupled hamiltonian ${\mathcal H}_\lambda$ keeps one and only one equilibrium configuration, at least whenever $\lambda$ is small enough. With respect to the direct sum of two single predator-prey models the full hamiltonian exhibits of course modifications both in the linear part, where the coefficients of the $q$ dependent terms are different (but this is really a minor modification, as those coefficients were arbitrary) and, more substantially, in the exponential part, where the arguments depend jointly upon $p$ and $q$ variables. From the theoretical point of view, one may of course resort to the KAM Theory (and theorem), having to do with hamiltonian perturbations.
Both hamiltonians being completely integrable on their own right, deciding which one to take as a perturbation of the other is largely a matter of taste or of phenomenological motivations, although, in view of the previous observation on the equilibrium configuration, keeping ${\mathcal H}_0$ as the unperturbed term seems a natural choice. At a first glance, when looking at the "full" Hamiltonian $${\mathcal H}_\lambda = {\mathcal H}_0 + \lambda {\mathcal H}_v$$ one may get a bit confused, in view of the fact that ${\mathcal H}_v$, in suitable canonical variables, takes the form of a one-body hamiltonian as well. However, this does not entail that we are considering just a linear combination of one body systems, because the coordinates that separate 
${\mathcal H}_v$ are not the same as those working for ${\mathcal H}_0$!

\noindent
When $\lambda = 0$ we deal with a completely integrable Hamiltonian system with 2 degrees of freedom, already separated in the sum of two commuting hamiltonians, whose level surfaces are compact (closed and bounded). So, in principle we have action-angle variables. But to our knowledge, in a generic case,  there is no way to get an explicit expression for the frequencies, even in the case of a single degree of freedom, unless we resort to numerical computations.
Let us now go back to the Hamiltonian ${\mathcal H}_0$, rewritten in terms of the variables $\xi$ and $\zeta$, that however, inspite of the possibly arising confusion, we denote again by $x$ and $y$, to simplify notations. Please remark that, being translated with respect to the original ones, those variables are no more constrained to the first quadrant, meaning that they can have both positive and negative values. In particular, the equilibrium position is now moved to the origin. So, we take as Hamiltonian function $ {\mathcal H}= af(x)+bf(y)$, whose level surfaces are given by the curve $af(x) + bf(y) = E$, and have the Hamilton's equation:

\begin{equation}
\dot x =b f^\prime (y) = b(1-\exp(y)); \quad \dot y= -af^\prime (x)=-a (\exp(x)-1) \label{newham}
\end{equation}

\subsection{About the period of the simplest model}
To get the period of the bounded motion described by (\ref{newham}), we write (for instance) the first of the above equations as
\begin{equation}
dt/dx= \frac{1}{1-\exp(y)}
\end{equation}
entailing that we have  to express $\exp(y)-1$ as a function of $E$ and $x$, taking into account that
\begin{equation}
	E = af(x) + bf(y); \quad f(x) = exp(x) - x - 1
\end{equation}
Now $x(y)$ varies on the whole real line, and correspondingly $f(x)(f(y))$ decreases monotonically (from $+\infty$ to $0$) on the negative semiline, vanishes at the origin and increases monotonically to $+\infty$ on the positive semiline. Hence for any positive $E$ both the functions $f(x)$ and $f(y)$ intersect the horizontal line $E$ = $const$ in two points, a negative one and a positive one, implying that they are only piecewise invertible. Taking care of such warning, we write
\begin{equation}
y = f^{-1}((E - af(x))/b)	
\end{equation}

\noindent
whence
\begin{equation}
dt= \frac {dx}{b(1-\exp(f^{-1}((E-af(x))/b)} 
\end{equation}
In order to get the period, (2.41) has to be integrated from two subsequent critical points
which are the negative and positive root of
\begin{equation}
	E/a = f(x)
\end{equation}	
In the literature one can find several different approaches to the problem of determining the period of the Volterra-Lotka system. A rich (however, possibly not exhausting) survey can be found in \cite{5}. There, 4 different methods are described, the first of them being due to Volterra himself. Here, we want to spend a few words on the one proposed by F. Rothe \cite{6}. In that paper, the author writes the system in Hamiltonian form (the same as we have introduced above)
for the Hamilton's function	
\begin{equation}
{\mathcal H}(x,y) = af(x) + bf(y)\label{pippa}
\end{equation}
\noindent
With	
$$f(x)) = \exp(x) - x -1$$
Rothe associates with (\ref{pippa} )the partition function
\begin{equation}
{\mathcal Z}(\beta) = \int_{-\infty}^{+\infty}	dx \int_{-\infty}^{+\infty} dy \exp(-\beta {\mathcal H}(x,y) )\label{part}
\end{equation}

\noindent	
which turns out to be expressible in terms of the Euler's $\Gamma$ function  as ${\mathcal Z}(\beta) = z(a \beta)z(b\beta)$, where
\begin{equation*}
z(\gamma)=\int_{-\infty}^{\infty}dx \exp[-\gamma(\exp(x)-x-1)] =\exp(\gamma)\gamma^{-\gamma}\Gamma(\gamma)
\end{equation*}

\noindent
On the other hand,  under the ergodic hypothesis, the partition function can be considered as the expectation value of the energy-period function $T(E)$, through :
\begin{equation}
 {\mathcal Z}(\beta) = \int_0^\infty dE T(E) \exp(-\beta E)
 \end{equation}
entailing that the energy-period function $T(E)$ is the inverse-Laplace transform of the partition function, and consequently is given by the following convolution:
\begin{equation}	
 T(E) = {\mathcal L}^{-1}[({\mathcal Z}(\beta)] ={\mathcal L}^{-1} [z](\beta a) \star{\mathcal L}^{-1}[z](\beta b)
 \end{equation}
To determine the inverse Laplace transform of $z$, let us consider again  $f(x) = \exp(x) - x - 1$, a real-analytic function mapping ${\mathbb R}$ to ${\mathbb R}_+$, monotonically decreasing (increasing) in the negative (positive) semiline, and denote by $x_- (x_+)$ the corresponding roots of the equation $f(x) = E$. Define now $\tau_+(E)= 1/|f^\prime (x_+)|=\frac{1}{E+x_+}$ and 
and similarly $ \tau_-(E) = 1/|f^\prime (x_-)| =\frac{1}{-E+|x_-|}$ and set $\tau(E) = \tau_+ + \tau_-$. Then:
 
 $$ z(\beta a) =  {\mathcal L}(a^{-1} \tau (E/a)); \quad  z(\beta b) =  {\mathcal L}(b^{-1} \tau (E/b))$$
 
 \noindent
Whence it follows the result: {\emph {The period of oscillations of the Volterra-Lotka system, parametrized as before, reads}}:
 \begin{equation}
 T(E)= \frac{1}{ab}\int_0^E ds \tau (s/b)\tau ((E-s)/a)
 \end{equation} 
On the other hand, we know from the textbooks on Hamiltonian mechanics (see for instance \cite{goldstein}), that in the case of a compact energy surface (meaning that all orbits are closed and bounded in the phase space) the frequencies, which are defined as the derivatives of the members of a set of commuting invariants with respect to the action integrals, coincide with the inverse of the corresponding periods. Here however, a problem arises. In a system with $N$ degrees of freedom (take for simplicity a completely integrable one), there are, say, $N$ (commuting)  integrals of motion $H_j$ and $N$ action integrals $J_k$. Moreover, in general, the integrals of motion will depend on several action integrals, so that typically we have matrices. The frequencies  are defined as the partial derivatives
 
 \begin{equation}
 \nu_j^{(k)}= \frac{\partial H_k}{\partial J_j}
 \end{equation}
The quantities $\frac{\partial J_k}{\partial H_j} $ will be the entries of the inverse matrix, which, unless everything is diagonal, won't be of course just  the inverse of the frequencies. So, a linear algebra operation, namely the inversion of a  matrix, is needed to recover the frequencies in the case we are given the ``action integrals" in terms of the ``constants of motion".
However, if we restrict for a moment our attention to the simplest case, governed by the hamiltonian ${\mathcal H}_0$ , the constants of motion are nothing but the single pair hamiltonians 
${\mathcal H}_1 = E_1$ and ${\mathcal H}_2 = E_2$. The equation defining the action is pretty similar to the one we have already encountered to calculate the period.
Indeed, the expression of $y$ as a function of $x$ and of the constants of motion has been derived in (2.40), which entails:
 \begin{equation}
 J_k= \oint dx_k(f^{-1}[(E_k-a_kf(x_k))/b_k]
 \end{equation}
 
 \noindent
whence:
\begin{equation}
 \nu_k = \oint dx_k \frac{b_k}{-1+\exp[f^{-1}(E_k-a_kf(x_k))/b_k]}
 \end{equation}
\medskip
\noindent 
 We emphasize again that we are considering the separated situation were ${\mathcal H}_1 = E_1$ and ${\mathcal H}_2 = E_2$, so that in the inversion procedure needed to express the linear momenta momenta  $y_k$ in terms of the conjugate coordinate $x_k$  and of the corresponding energy level, we can attach to each degree of freedom it's own energy value. In other words, the action integral $J_k$ depends just upon $E_k$, so that the matrix introduced above is diagonal and the frequency $\nu_k^{(k)}$  is the reciprocal of the (total) derivative  $\frac {d E_k}{dJ_k}$.
Moreover, the procedure introduced in \cite{6} to calculate the period in the original Lotka-Volterra has a trivial extension to the case of the direct sum of two systems. The partition function of the direct sum is in fact the product of those pertaining to each of them. Indeed, in that case the partition function will be given by
\begin{equation}
{\mathcal Z}^{(2)}(\beta) = \Pi_{i=1}^2\int_{-\infty}^\infty dx_i \int_{-\infty}^\infty dy_i \exp(-\beta {\mathcal H}_i(x_i,y_i)
\end{equation}
the functions ${\mathcal H}_i$ being the same for both systems, up to possible changes of the coefficients.

\section{Concluding remarks}

We stress that the  results reported above are not complete. First of all, even in the completely integrable case, we lack an explicit reduction to quadratures of the equations of motion for an arbitrary $N$: here, we have performed it only up to two predator-pray pairs. Other issues, ubiquitous in the theory of integrable systems, such as the existence of a Lax representation and of a discrete (in time)  integrable version of the continuous dynamics are at the moment out of our description. 
A further question, more important in view of the possible applications, is related to the construction of a richer model, still enjoying the complete integrability property, but involving a larger number of conjugated pairs of variables, all of them being significant in the economic-ecological approach. Work is progress in all the outlined directions. 






\vfill

\end{document}